\documentclass[aps,pra,twocolumn]{revtex4}
\usepackage{graphicx}

\usepackage{bm}

\begin{document}

\title{Signatures of non-locality in the first-order coherence of
the scattered light} \author{Priscilla Ca\~nizares,$^1$ Tobias
G\"orler,$^2$ Juan Pablo Paz,$^3$ Giovanna Morigi,$^1$ and Wolfgang Schleich$^2$}
\affiliation{$^1$ Departament de Fisica, Universitat Autonoma de Barcelona, E-08193 Bellaterra, Spain,\\
$^2$ Abteilung Quantenphysik, Universit\"at Ulm, D-89069 Ulm, Germany,\\
$^3$ Departamento de fisica, Universidad de Buenos Aires, 1428 Buenos
Aires, Argentina.}

\date{\today} \begin{abstract} The spatial coherence
of an atomic wavepacket can be detected in the scattered photons,
even when the center-of-mass motion is in the quantum coherent
superposition of two distant, non-overlapping wave packets. Spatial coherence manifests itself in the power spectrum of the emitted photons, whose spectral components can exhibit interference fringes as a function of the emission angle. The contrast and the phase of this  interference pattern provide information about the quantum state of the center of mass of the scattering atom.
\end{abstract} \maketitle

\section{Introduction} Non-locality is a quantum mechanical
property, which has no classical counterpart. Its disappearance in
the classical world is related to the interaction with the
surrounding environment, which in several cases can be identified
with the electromagnetic field~\cite{Decoherence:1}. In fact,
photon scattering may have the same effect as a projective
measurement of the position of the system inducing the effective
collapse of the quantum state into a position eigenstate, thus
destroying spatial coherence \cite{Decoherence}. Nevertheless, one
can identify other physical situations where photon scattering
enforces spatial coherence. Laser cooling is a suggestive example,
where quantum states of the atomic center--of--mass
motion are prepared at the steady state of a process based on
photon scattering, achieving coherence lengths which can exceed
the laser wavelength~\cite{Cohen-Tannoudij91}. The emitted photons
carry the information about the atomic state, whose dynamics can
be characterized, for instance, in the correlation functions of
the light~\cite{Levy,ResFluo}. This observation leads to the natural expectation that non--locality can be measured by means of photon
scattering. Even if in general the photons will destroy
spatial coherence, a quantum scatterer will imprint 
features of its state in the photons emitted during the transient
dynamics. 

\begin{figure}[!th] \includegraphics[width=6cm]{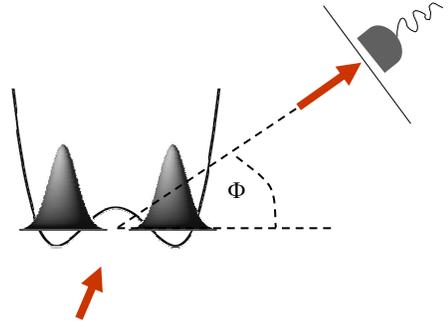}
\caption{An atom is confined by double--well potential and is driven by a laser. The intensity of the scattered light is measured in the far field by a detector which is sensitive to intensity gradients over the emission angle $\Phi$. The spectral components of the resonance fluorescence, scattered by the atom in the ground state of the potential, exhibit interference fringes as a function of $\Phi$, whose phase and contrast are determined by the coherence properties of the atomic wave packet. }
\label{Fig:1} \end{figure}

In this article we analyze light scattering by a particle prepared in the
ground state of a double--well potential, and study the
first--order coherence properties of the scattered photons as a
function of the distance between the wells and of the size of the
atomic wave packet, when the atomic transition is driven by a weak
laser pulse. The setup is sketched in Fig.~\ref{Fig:1},
and is reminescent to a Young
(double--slit) interference experiment. However, the slit is here
a single atom, which is prepared in a coherent superposition of
two locations. We show that in general non--locality can be
measured when the particle can tunnel between the two wells, and
it manifests itself in the spatial periodic modulation of the
elastic peak and of the Stokes and anti-Stokes signal. Moreover,
we predict that in certain situations it is possible to detect the
phase of the quantum superposition between two spatially distant
wave-packets. These results are discussed in connection to recent
theoretical works, which studied matter wave scattering by a
quantum object~\cite{Beenakker02,Folman06}.

\section{Coherence of light from a quantum scatterer}

We consider an atom of mass $M$, which is prepared in the stable electronic
state $|g\rangle$ and whose center-of-mass is confined by a double--well potential $V_g({\bf x})$. The center of mass is in a
stable state $|\psi\rangle$, eigenstate of the free Hamiltonian
\begin{eqnarray} H_{\rm mec}^{(g)} =\frac{{\bf p}^2}{2M}+V_g({\bf x})
\end{eqnarray} where ${\bf x},{\bf p}$ are position and momentum
of the atom. State $|\psi\rangle$ is described by the
superposition of the ground states of each wells
$|\psi_{\pm}\rangle$, \begin{equation} \label{State:0}
|\psi\rangle={\cal N}\left(\cos\theta|\psi_-\rangle+{\rm
e}^{i\phi}\sin\theta|\psi_+\rangle\right)\end{equation} where
$\theta,\phi$ are the azimuthal and polar angles and ${\cal N}$
gives the proper normalization. The distance between the wells
centers is $d=|{\bf d}|$, and the potential of each well is
approximated by two harmonic oscillators at frequency $\nu$ and
ground-state wave packets \begin{equation}
\label{toy-wf}\langle{\bf x}|\psi_{\pm}\rangle=\frac{{\rm
e}^{-({\bf x}\mp{\bf d}/2)^2/(4a_0^2)}}{(2\pi a_0^2)^{3/4}}
\label{psi+-}\end{equation} with $a_0=\sqrt{\hbar/2M\nu}$ size of
ground state of each oscillator. A laser pulse, centered at
frequency $\omega_L$ couples to the dipole transition
$|g\rangle\to |e\rangle$ at frequency $\omega_0$, and the
corresponding dynamics is described by the term \begin{eqnarray}
W_L=\hbar\Omega\int {\rm d}\omega
f(\omega)\left(\sigma^{\dagger}{\rm e}^{-{\rm i}(\omega t-{\bf
k(\omega)}\cdot {\bf x})}+{\rm H.c.}\right)\end{eqnarray} with
$\sigma=|g\rangle\langle e|$ and $\sigma^{\dagger}$ its adjoint,
$f(\omega)$ is the normalized spectral distribution over the
frequencies, centered in $\omega_L$ and with width $\Delta\omega$,
${\bf k(\omega)}$ the wave vector at frequency $\omega$, and
$\Omega$ is the Rabi frequency at frequency $\omega_L$. The
excited state couples to states $|g\rangle$ and
$|g^{\prime}\rangle$  with strength $g_{\bf k}$ and $g_{\bf
k}^{\prime}$, emitting a photon into the mode $\omega_{\bf k}$ and
$\omega_{\bf k}^{\prime}$ with wavevector ${\bf k}$ and ${\bf
k^{\prime}}$, respectively. The scattered photons are collected in
the far field by a detector, which is sensitive to the gradient of
the scattered intensity over the angle of emission $\Phi$ and to
the frequency $\omega$, hence to the wavevector ${\bf k}$. In the
far field the spherical waves of the scattered fields are well
approximated by plane waves and the scattering cross section can
be put in direct correspondence with the intensity measured at the
detector~\cite{WinelandYoung}. We thus evaluate the scattering
rate $R_{\bf k}$ of photons with wave vector ${\bf k}$, and thus
into the solid angle defined by the direction of ${\bf k}$, and
correspondingly the scattering rate $R_{\bf k}^{\prime}$. These
can be measured separately, for instance by using a polarizing
filter before the detector.

Let us consider the rate of photon scattering along the transition
$|g\rangle\to|e\rangle$. The scattering rate over a time $T$ is
$R_{\bf k}=P_{\bf k}/T$, where $P_{\bf k}=Tg_{\bf k}^2\Omega^2\int
{\rm d}\omega\int {\rm
d}\omega^{\prime}f(\omega)f(\omega^{\prime}){\cal A}$ is the
probability for a photon of being emitted in the mode with wave
vector ${\bf k}$, and \begin{eqnarray} {\cal A}
&=&\frac{1}{T}\int_0^t{\rm d}\tau\int_0^{\tau}{\rm d}\tau^{\prime} \int_0^t{\rm d}\bar{\tau}\int_0^{\tau}{\rm d}\bar{\tau}^{\prime}{\rm e}^{{\rm i}(\omega_0-\omega_{\bf k})\bar{\tau}}{\rm e}^{-{\rm i}(\omega_0-\omega)\bar{\tau}^{\prime}}\nonumber\\
& &\times {\rm e}^{-{\rm i}(\omega_0-\omega_{\bf k})\tau} {\rm
e}^{{\rm i}(\omega_0-\omega^{\prime})\tau^{\prime}}{\cal
W}(\tau,\tau^{\prime},\bar{\tau},\bar{\tau}^{\prime})\label{Prob}
\end{eqnarray} where the initial state is the atom in $|g,\psi\rangle$, Eq.~(\ref{State:0}), and the electromagnetic field in
the vacuum. In writing Eq.~(\ref{Prob}) we have introduced ${\cal
W}=\langle\psi|\bar{U}^{\dagger}U|\psi\rangle$ with
\begin{eqnarray} U={\rm e}^{{\rm i}H_{\rm
mec}^{(g)}\tau/\hbar}{\rm e}^{-{\rm i}{\bf k}\cdot {\bf x}} {\rm
e}^{-{\rm i}H_{\rm mec}^{(e)}(\tau-\tau^{\prime})/\hbar}{\rm
e}^{{\rm i}{\bf k_L}\cdot {\bf x}}{\rm e}^{-{\rm
i}H_{\rm mec}^{(g)}\tau^{\prime}/\hbar}\nonumber\\
\label{U} \end{eqnarray} and
$\bar{U}=U|_{\tau\to\bar{\tau};\tau^{\prime}\to\bar{\tau}^{\prime}},$
where $H_{\rm mec}^{(e)}$ is the Hamiltonian for the atomic center
of mass in state $|e\rangle$. The rate of Raman scattering, $R_{\bf
k}^{\prime}$, is analogously defined, where we assume that the
Hamiltonian of the final states is $H_{\rm mec}^{(g')}={\bf
p}^2/2M$, namely the atom propagates freely. This situation may be
realized with magnetic traps, where $|g\rangle$ is a weak--field
seeker and $|g^{\prime}\rangle$ is a state of the hyperfine multiplet
with zero magnetic moment~\cite{Schumm}. The dependence on the dipole pattern of emission ${\cal D}(\Phi)$ in the
solid angle $\Phi$ enters in the rates through the factor $g_{\bf
k}^2\propto \gamma {\cal D}(\Phi)$, with $\gamma$ linewidth of
state $|e\rangle$. In the far field we neglect the variation of
this factor over the angle.

Substituting the explicit form of $|\psi\rangle$ into ${\cal W}$,
one can write the transition amplitude as the sum of four
contributions, \begin{eqnarray} \label{W}{\cal W} ={\cal
N}^2\left[ \sin^2\theta\langle
\psi_+|\bar{U}^{\dagger}U|\psi_+\rangle+\cos^2\theta\langle\psi_{-}|\bar{U}^{\dagger}U|\psi_{-}\rangle
\right.\\
\nonumber\left. +\sin\theta\cos\theta\left( {\rm e}^{-{\rm
i}\phi}\langle\psi_+|\bar{U}^{\dagger}U|\psi_-\rangle + {\rm
e}^{{\rm
i}\phi}\langle\psi_-|\bar{U}^{\dagger}U|\psi_+\rangle\right)\right]\nonumber
\end{eqnarray} where the last two terms are due to the non-local
properties of the initial state. Clearly, the contribution of the
latter is maximal with respect to $\theta$ when $\theta=\pi/4$,
i.e., when left and right states are initially equally
populated. We thus restrict to this case, and identify situations
where these terms do not vanish, allowing to observe features due
to the non-local properties of the atomic wave packet in the
scattered light.

In the following we evalute scattering by an atom driven by a weak laser pulse, when the pulse duration is such that one can neglect the width $\Delta\omega$ about the mean value $\omega_L$. We consider the case in which the laser is
far-off resonance from the dipole transition
$|g\rangle\to|e\rangle$, such that state $|e\rangle$ is
practically empty. In this regime we can neglect the propagator in
the excited state in Eq.~(\ref{U}), as one is on the flat
tail of the Lorentz curve of the atomic resonance. We focus on the evaluation of
the rates of Rayleigh scattering, when the final
internal state of the atom is equal to the initial state
$|g\rangle$, and of Raman scattering, when the atom is in the
state $|g^{\prime}\rangle$ after scattering one photon. We note that for large laser detuning and short interaction times we can
approximate $U(\tau,\tau^{\prime})\approx U(0,0)$ in
Eq.~(\ref{U}), and ${\cal W}\approx 1$: The scattering rate is the same as
the atom was a point-like scatterer. In fact, for short interaction times the motion is essentially classical. We thus focus on long interaction times, so to be able to resolve the spectral components of the resonance fluorescence.

\subsection{Rayleigh scattering}

We now evaluate the rate of Rayleigh scattering $R_{\bf k}$ when the atom has been prepared in the symmetric state of the
double-well potential, and consider the photons which are elastically scattered and inelastically scattered to the antisymmetric state. Be $\delta\nu$ the frequency splitting between the two
states. From Eq.~(\ref{toy-wf}), term ${\cal W}$ is explicitly
evaluated using the relation ${\rm e}^{{\rm i}({\bf k_L}-{\bf
k})\cdot{\bf x}}|\psi_{\pm}\rangle =|\beta_{\pm}\rangle{\rm
e}^{\pm{\rm i}({\bf k_L}-{\bf k})\cdot{\bf d}/4}$ where
$|\beta_{\pm}\rangle$ is a coherent state with amplitude
$\beta_{\pm}=\pm{\bf d}/4a_0+{\rm i}a_0({\bf k_L}-{\bf k})$. With
these relations, \begin{widetext}\begin{eqnarray} \label{W:F:1}
{\cal A}\approx \frac{1}{\Delta^2}{\rm e}^{-|\Delta{\mathbf
k}|^2a_0^2}\left[\frac{1}{(1+\epsilon)^2}\left(\cos\left(\frac{\Delta{\mathbf
k}\cdot{\bf
d}}{2}\right)+\epsilon\right)^2\delta^{(T)}(\omega_{\bf
k}-\omega_L)+\frac{1}{1-\epsilon^2}\sin^2\left(\frac{\Delta{\mathbf
k}\cdot{\bf d}}{2}\right)\delta^{(T)}(\omega_{\bf
k}-\omega_L+\delta\nu)\right] \end{eqnarray} \end{widetext}
where $\Delta=\omega_L-\omega_0$, $\Delta{\bf k}={\bf k_L}-{\bf
k}$ is the difference between the wavevectors of the incoming and
outcoming photons, $\delta^{(T)}(\omega)$ is the diffraction
function~\cite{AtomPhoton}, and \begin{equation} \epsilon={\rm
e}^{-d^2/8a_0^2} \end{equation} is the overlap between the two
wavepackets. The finite width $\Delta\omega$ of the incident laser
pulse can be neglected for $\Delta\omega\ll 2\pi c/d$. Moreover,
if $\Delta\omega\ll\delta\nu$, then $R_{\bf k}\propto {\cal A}$.
This result shows that the elastic peak, at $\omega_{\bf
k}=\omega_L$, and the Stokes sideband, at $\omega_{\bf
k}=\omega_L-\delta\nu$ exhibit interference fringes on the screen as a function of the emission angle,
which are in opposition of phase and have periodicity determined
by the distance $d$ between the wells. When the antisymmetric
state is initially occupied, one observes the signal at the anti-Stokes
frequency, at $\omega_{\bf k}=\omega_L+\delta\nu$, which oscillates in
phase with the Stokes sideband. These sinusoidal signals have Gaussian envelope
with width $1/a_0$, thus determined by the size of the wave
packets localized at each well. Hence, the number of observed
fringes is small, since ideally $d\sim a_0$. In absence of a spectral filter,
resolving the splitting between the doublet, the scattered intensity still depends on the solid angle, with
a visibility determined by $\epsilon$, which approaches unity as
the spatial overlap between the two wave packets increases. 

In a time-picture, the spectral resolution of elastic and Stokes components corresponds to selecting interaction times
$T$ which are larger than the tunnelling time
$1/\delta\nu$, such that during the interaction with the photon the atom tunnels between the wells. Hence, the system is analogous to a
Young's interference setup with one single slit, which is in a coherent
superposition of two spatial locations. Remarkably,
Eq.~(\ref{W:F:1}) also predicts an interference signal when the
distance between the wells is very large, $\epsilon\to 0$, and the
tunneling rate thus vanishes, provided that elastic and Stokes sideband are resolved in the spectrum of resonance fluorescence. However, in this regime the needed spectral resolution scales as $\delta\nu$, and thus is extremely
small, requiring integration over diverging time intervals $T$. 

Equation~(\ref{W:F:1}) takes into account
the finite size of the atomic wave packet and the mechanical
effect of the scattered photon on the atom. 
This result is the photonic counterpart of the scattering rate for
matter waves by a quantum object derived
in~\cite{Beenakker02,Folman06}. The scattering rate
in~\cite{Beenakker02}, which was obtained for a point-like
scatterer, is recovered taking $\epsilon\to 0$. In
Ref.~\cite{Folman06} the interference pattern is explained in
terms of entanglement swapping of the state of the scatterer with
the state of the scattered matter wave. This
interpretation applies also to photon scattering: the emerging
photon is entangled with the scattering
atom~\cite{Eberly,MonroeWeinfurter}, and interference between the
paths of photon scattering through each well is hence possible
when the two states, $|\psi_+\rangle$ and $|\psi_-\rangle$, have
finite overlap. The interaction with the atom prepared in a superposition of two orthogonal wave-packets
imprints a which-way information on the photon~\cite{Duerr98}, and one can put the finite overlap between the wave-packets in
direct connection with the fringe visibility at the screen~\cite{Englert96}.

\subsection{Raman scattering}

Is it then possible to detect by photon scattering whether the
atom is in a non-local state, if there is no spatial overlap
between the two wave packets in which the atom has been prepared?
Let us assume the atom is initially in state
$|g,\psi\rangle$ when $|{\bf d}|\gg a_0$, so that the tunneling
rate vanishes, $\epsilon\to 0$. This state can be prepared, for instance,
via adiabatic manipulation of the center-of-mass wave packet, as
described in~\cite{Immanuel03}. We assume now that the internal
state is coupled by Raman scattering to the internal state
$|g^{\prime}\rangle$, which is not trapped. In this case, a photon
of wave vector ${\bf k^{\prime}}$ is emitted in a spontaneous
Raman process, which is detected at the screen. Accordingly, the
atom in state $|g^{\prime}\rangle$ propagates freely with momentum
$\hbar({\bf k_L}-{\bf k^{\prime}})$. The corresponding signal at
the detector is given by \begin{widetext} \begin{eqnarray} {\cal
A}^{\prime}\propto
\frac{1}{\Delta^2}{\rm
e}^{-2|\Delta k|^2a_0^2} \left[1+\sin 2\theta \cos({\bf \Delta
k}\cdot {\bf d}+\phi)\right] \delta^{(T)}(\omega_L-\omega_{\bf
k}+\delta\omega_{g'}+\hbar{\bf \Delta k}^2/2M)\label{A:1}\end{eqnarray}
\end{widetext} where ${\bf \Delta k}={\bf k_L}-{\bf k^{\prime}}$ and $\delta\omega_{g'}$ accounts for the frequency difference between initial and final state.
For $\Delta\omega\ll 2\pi c/d,\hbar{\bf \Delta k}^2/2M$, then
$R_{\bf k}^{\prime}\propto {\cal A}^{\prime}$ and the intensity at
the detector exhibits an interference signal, provided that the
frequency shift due to photon recoil is spectrally resolved. This
signal has periodicity determined by $d$, it has maximum contrast
when the occupation probability of the two wells is equal,
$\theta=\pi/4$, and it depends on the phase $\phi$ of the
initial state, Eq.~(\ref{State:0}). The situation here
considered is similar to the experimental setup realized in
Ref.~\cite{Saba05}, which was used for detecting the phase between
two Bose-Einstein condensates by measuring interference in the flux of atoms outcoupled by stimulated Bragg scattering. Equation~(\ref{A:1}) shows that interference is already present at the single-particle level.
As observed in~\cite{Saba05}, this effect is found whenever the
final state has finite overlap with both initial wave packets. In
analogy with interferometry, the final state is a quantum eraser,
which projects the two orthogonal wave packets into the same
state, thus restoring interference~\cite{ScullyDruhl}. The detection efficiency of the scattered photons in this kind of experiment
can be improved using stimulated Raman scattering or enhancing the rate of photon emission by means of a resonator, in a setup like the one discussed in Ref.~\cite{Keller}. We remark that a signal similar to the one in Eq.~(\ref{A:1}) is found also in the photons scattered by an atomic dipole, when the atomic wave packet in the excited state $|e\rangle$ has finite overlap with both wells and the vibrational excitations of $|e\rangle$ are spectrally resolved. In this case, propagation in the excited state leads to an overlap between the two initial wave packets and acts as a quantum eraser. Hence, inelastic scattering restores the visibility of the fringes. This is a remarkably different behaviour from the one studied in~\cite{WinelandYoung}, where the coherence of light scattered by two atoms is destroyed by saturation effects.

\section{Conclusions}

Using simple, but experimentally plausible models we have shown
that the features associated with non-locality can be measured in
the scattered light. The interaction of a photon with an atom in a double--well potential can be thought of as
a photon interferometer, and coherence between the two interaction paths appear through spatial modulation in the
signals of the spectrum of resonance fluorescence. This setup can be generalized to the case of quantum transport of an atom in a periodic potential, such as an optical lattice: Photon scattering will act as a multipath interferometer, providing information on the coherence properties of atomic motion. Such setup could be used in order to study the coherence length of a reservoir coupling to the atom, which could be, for instance, an atomic gas~\cite{JPaz}.

These results give a further example of how the spatial coherence of matter waves appear
in the coherence properties of the scattered light, and may be of relevance for the realization
of quantum light sources~\cite{IonTrap,Keller}.\\

\acknowledgements
Many of us have often heard Herbert Walther express his amusement of and
quiet admiration for a close student-teacher relationship by quoting
Friedrich Schiller's Wallenstein: "Wie er r\"auspert und wie er spuckt,
das habt ihr ihm gl\"ucklich abgeguckt" (The way he clears his throat, the
way he spits, you have emulated very well from him). Most likely Herbert
Walther did not realize how much more he himself was a role model as a
teacher for all of us. We were the students who tried to become a
scientist like him. His excitement about a new discovery, his love of
and dedication to science and most importantly his abundant energy has
infected all of us. It is hard to believe that our hero has left us; he
still could have taught us so much more. It has been a great privilege
for us to have worked with him and we dedicate the present paper to his
memory hoping that it would have found his interest and approval.

G.M. acknowledges discussions with E. Demler, J. Eschner, S. Rist, and U. Weiss, and the kind
hospitality of the Los Alamos National Laboratory during
completion of this work. Support by the European Commission
(CONQUEST, MRTN-CT-2003-505089; EMALI, MRTN-CT-2006-035369), the
scientific exchange programme Germany-Spain (HA2005-0001 and
D/05/50582), the Spanish Ministerio de Educaci\'on y Ciencia
(Ramon-y-Cajal; QLIQS, FIS2005-08257; Consolider Ingenio 2010
QOIT, CSD2006-00019), and the scientific exchange programme
Spain-Argentina (AECI A/3448/05) are acknowledged.


\begin{thebibliography}{99} \bibitem{Decoherence:1}
E. Joos, H.D. Zeh, C. Kiefer, D. Giulini, J. Kupsch, and I.-O. Stamatescu,
{\it Decoherence and the Appearance of a Classical World in Quantum Theory},
Springer Verlag (Heidelberg, 2003).
\bibitem{Decoherence} W.H. Zurek, Phys. Today {\bf 44}, 36 (1991);
J. P. Paz and W. H. Zurek, in {\it Coherent matter waves, Les Houches
Session LXXII}, R Kaiser, C Westbrook and F David eds., EDP Sciences
(Springer Verlag, Berlin, 2001), pp. 533-614.
\bibitem{Cohen-Tannoudij91}
C. Cohen-Tannoudij, F. Bardou, and A. Aspect, in {\it Laser Spectroscopy X}, eds. M. Ducloy, E. Giacobino, and G. Camy (World Scientific, Singapore, 1992), pp. 3-14.
\bibitem{Levy}
H. Katori, S. Schlipf, and H. Walther,
Phys. Rev. Lett. {\bf 79}, 2221 (1997).
\bibitem{ResFluo}
J.T. Hoffges, H.W. Baldauf, T. Eichler, S.R. Helmfrid, and H. Walther,
Advances in Quantum Chemistry {\bf 30}, 65 (1998);
Ch. Raab, J. Eschner, J. Bolle, H. Oberst, F. Schmidt-Kaler, and R. Blatt,
Phys. Rev. Lett. {\bf 85}, 538 (2000).
\bibitem{Beenakker02}
H. Schomerus, Y. Noat, J. Dalibard, and C.W.J. Beenakker,
Europhys. Lett. {\bf 57}, 651 (2002).
\bibitem{Folman06}
D. Rohrlich, Y. Neiman, Y. Japha, and R. Folman, Phys. Rev. Lett.
{\bf 96}, 173601 (2006).
\bibitem{WinelandYoung}
U. Eichmann, J.C. Bergquist, J.J. Bollinger, J.M. Gilligan, W.M.
Itano, D.J. Wineland, and M.G. Raizen, Phys. Rev. Lett. {\bf 70},
2359 (1993);
W.M. Itano, J.C. Bergquist, J.J. Bollinger, D.J. Wineland, U.
Eichmann, and M.G. Raizen, Phys. Rev. A {\bf 57}, 4176 (1998).
\bibitem{Schumm}
T. Schumm, S. Hofferberth, L.M. Andersson, S. Wildermuth, S. Groth, I. Bar-Joseph, J. Schmiedmayer, P. Kr\"uger,
Nature Physics {\bf 1}, 57 (2005).
\bibitem{AtomPhoton}
C. Cohen-Tannoudij, J. Dupont-Roc, and G. Grynberg, {\it Atom-Photon Interactions}, Wiley ed. (New York, 1992).
\bibitem{Eberly}
K. W. Chan, C. K. Law, and J. H. Eberly, Phys. Rev. Lett. {\bf
88}, 100402 (2002);
K. W. Chan, C. K. Law, and J. H. Eberly, Phys. Rev. A {\bf 68},
022110 (2003)
\bibitem{MonroeWeinfurter}
B. Blinov, D. L. Moehring, L.-M. Duan, and C. Monroe, Nature
(London) {\bf 428}, 153 (2004);
J.\ Volz, M.\ Weber, D.\ Schlenk, W.\ Rosenfeld, J.\ Vrana, K.\
Saucke, C.\ Kurtsiefer, H.\ Weinfurter, Phys. Rev. Lett. {\bf 96}, 030404 (2006).
\bibitem{Duerr98}
S. D\"urr, T. Nonn, and G.Rempe, Nature (London) {\bf 395}, 33
(1998);  Phys. Rev. Lett. {\bf 81}, 5705 (1998).
\bibitem{Englert96}
B.-G. Englert, Phys. Rev. Lett. {\bf 77}, 2154 (1996).
\bibitem{Immanuel03}
O. Mandel, M. Greiner, A. Widera, T. Rom, T. W. H\"ansch, and I.
Bloch, Phys. Rev. Lett. {\bf 91}, 010407 (2003).
\bibitem{Saba05}
M. Saba, T.A. Pasquini, C. Sanner, Y. Shin. W. Ketterle, and D.E. Pritchard, Science {\bf 307}, 1945 (2005).
\bibitem{ScullyDruhl}
M.O. Scully and K. Dr\"uhl, Phys. Rev. A {\bf 25}, 2208 (1982).
\bibitem{Keller}
M. Keller, B. Lange, K. Hayasaka, W. Lange, and H. Walther, Nature
{\bf 431}, 1075 (2004).
\bibitem{JPaz} M.R. Gallis and G.N. Fleming, Phys. Rev. A \textbf{42}, 38 (1990);
K. Horneberger, Phys. Rev. Lett. \textbf{97}, 060601 (2006).
\bibitem{IonTrap}
G.M. Meyer, H.J. Briegel, and H. Walther, Europhys. Lett. {\bf 37}, 317 (1997).
\end{thebibliography}
\end{document}